\documentclass[12pt]{article}

\usepackage[T1]{fontenc} 
\usepackage{array} 
\usepackage{amsmath}
\usepackage{amssymb}
\usepackage{latexsym}
\usepackage{epic}
\usepackage{eepic}
\usepackage{epsfig, color}
\usepackage{amsfonts}
\usepackage{amsbsy}
\usepackage{graphicx}
\usepackage{epsfig,ae}

\newtheorem{remark}{Remark} 
 
\newcommand{\mab}[1]{{\bf #1}}

\title{Manifold and metric in numerical solution of the quasi-static 
  electromagnetic boundary value problems}

\date{} 
\author{Pasi Raumonen\footnote{This work was supported by 
    Academy of Finland, project number 5211066.}, Saku Suuriniemi, 
  Timo Tarhasaari, \\ Lauri Kettunen\\
 {\small Tampere University of Technology, 
   Institute of Electromagnetics,} \\ 
 {\small  P.O. Box 692,  FIN-33101 Tampere, Finland.}}

\begin{document} 

\maketitle

\begin{abstract}
  Classical vector analysis is the predominant formalism used by 
  engineers of computational electromagnetism, despite the fact 
  that manifold as a theoretical concept has existed for 
  a century. This paper discusses the benefits of manifolds over 
  the traditional approach in practical problems of modelling. With 
  a structural approach, it outlines the role and interdependence 
  of coordinate systems, metric, constitutive equations, and fields, 
  and relates them to practical problems of quasi-static 
  computational electromagnetics: mesh generation, open-boundary 
  problems, and electromagnetic-mechanical coupled problems 
  involving motion and deformation. The proposed procedures also 
  imply improvements to the flexibility of the modelling software.
\end{abstract} 

%\keywords{Electromagnetics, Manifolds, Metric tensors, Boundary value
%  problems} 

\section{Introduction}

The objective of solver software systems for electromagnetic boundary 
value problems (BVP) is to solve an arbitrary problem from a specific 
problem class. However, this is not possible in practice: One of the 
restricting factors is the difficulty of the mesh generation. This is 
related to the fact that to pose a BVP, the user needs to define a 
geometry of the BVP domain. The geometry is described with a coordinate 
system and certain choices of coordinate systems may cause too much 
round-off error with floating point arithmetic.

Often the problem geometries lend themselves to a particular coordinate 
system, and therefore the software systems usually offer a list of 
coordinate systems to choose from. To offer a user interface with 
a list of coordinate systems, many structures in the software are 
fixed: For example, the software often has one intrinsic coordinate 
system, usually Cartesian, which it uses in all calculations. 
Furthermore, the basic material parameters, such as the permittivity 
of a vacuum, as well as the metric (inner product) in the intrinsic 
coordinate system are pre-decided. When the user gives the model's 
geometry, for example, in the spherical coordinates, the software 
maps them to the intrinsic Cartesian system. 

\begin{figure}
  \centering\includegraphics[width=\textwidth]{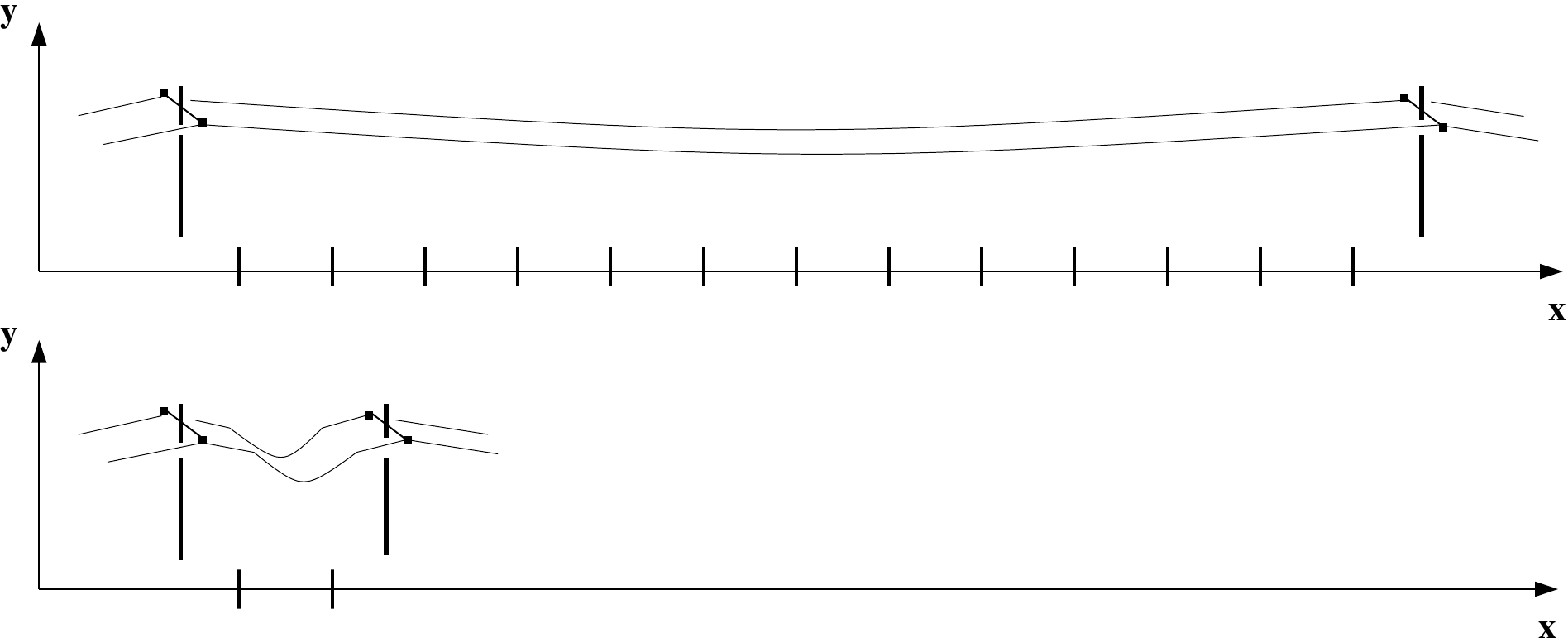}
  \caption{Power line in two coordinate system.}
  \label{fig:powerlines} 
\end{figure}

Frequently, the floating point problems \emph{leave the modeler 
  without any mesh at all} and prevent the solution of the BVP:
Practical computations with the intrinsic coordinate system 
requires that the coordinate numbers are represented with floating 
point numbers. This makes mesh generation particularly challenging 
for BVPs with details of the domain varying significantly in size.
For example, consider the power line on top of 
figure~\ref{fig:powerlines}. The thickness of the cables and the 
structures in the supports are much smaller in scale than the 
distance between the two supports. Meshes are difficult to 
generate because the floating point representation gives us 
labels to only a finite, although large, number of points. 
For example, as a floating point number grows in magnitude, the 
distance between points separable by numeric labels increases, 
and, at some point, round-off errors become big enough to make 
labels of distinct points coincide unintentionally. Generally, 
mesh generation is based on geometric predicates \cite{Shewchuk}, 
some of which are difficult to calculate with floating point 
numbers. Mesh generation becomes hard or even impossible with double 
precision floating point numbers when the scale variation, the 
ratio of the largest and the smallest dimensions, is in the order 
of $10^4$ or higher. This estimate for scale variation is based 
on the experiences with the BVP presented in the section 7.

The mesh generation problems arise from the way the BVP was described 
in a particular coordinate system. Is it possible to alleviate these 
problems? With the help of modern mathematics, manifolds and 
differential geometry, the role of mathematical structures on the 
description of a BVP can be outlined. In this paper we strongly 
promote a structural approach. Particularly, we will show that for 
a given BVP, the elements of the triplet \{coordinate system, metric, 
material parameters\} are interdependent, and the triplet constitutes 
an entity. In fact, there is a whole class of these triplets for 
each BVP: Each triplet describes a geometry of the BVP domain 
and the constitutive equations, which are needed to define an 
electromagnetic BVP. Specifically, the numerical values of material 
parameters are respect to given metric. For example, the numerical 
value of the permittivity of vacuum is different with respect to
meters than with respect to inches. Although the metric appears in 
the triplets, the electromagnetic theory essentially concerns the 
works related to displacements of particles and the total energy of 
systems. The role of metric is central only in the classical 
view---vector analysis---whereas the modern view of differential 
geometry reveals the independence of the electromagnetic theory from 
a particular metric. This has been known for long 
\cite{Kottler}\cite{Cartan}\cite{Dantzig}\cite{Schrödinger}\cite{Post}, 
but not that well recognized in engineering. Hence, 
\emph{the modeller may select the metric at will}. 
This is much greater a freedom in the context of manifold than 
a redefinition of the inner product in the single codomain vector 
space of a vector field in the classical view. We use the modern 
view to show how to formulate electromagnetic problems with the 
classical vector formalism in different metrics. 

From the software design point of view, there is no need to fix the 
whole triplet in a software system. Just like the software system 
is not intended to solve a particular problem, but any problem from 
a class of problems, the solver software should provide the 
possibility to use any triplet from the triplet class. It should 
not be even restricted to a single triplet, because manifold promotes 
local coordinate systems that provide further flexibility. This will 
help overcome floating point problems in the calculations of 
the geometric predicates and thus will enlarge the class of 
practically solvable problems: Our approach suggests giving up 
the metric and the coordinate system strongly suggested by human 
intuition. For example, if the mesh generation fails with the
coordinate system in the top of figure \ref{fig:powerlines},
one could use the coordinate system in the bottom of the same 
figure, when the calculations of the predicates are easier.
This time it is more likely that one gets a mesh and thus 
a numerical solution. However, there is usually a price to pay: 
numerical error can be large because the mesh quality may be poor.

The problems of mesh generation are just one example of practical 
challenges in numerics that can be engaged with tools provided 
by manifold. The problems with motion and deformations, such as 
the deformation effects of magnetostriction, can be modelled 
with a single mesh: motions and deformations can be modelled as 
changes in the metric. This can speed up the solution process
substantially. Finally, the solutions for open boundary problems,
that are solved by finite domains with so-called infinite elements 
in the boundary region, are an example of an application where 
a convenient triplet is chosen for solution.

\section{Structural Background}

We need many mathematical structures to pose and describe an 
electromagnetic BVP. The starting point of this description is 
a point set that is the BVP domain. The points of this set 
correspond to observations on distance measurements. Then 
we will add one structure at a time to this point set, such that 
finally we have all that is needed for the description.

\subsection{Metric Space}

At first we need to specify a point set for the BVP domain. The 
points in the domain correspond to the points observed in reality,
in the following sense: The observed points are considered distinct 
if and only if a nonzero distance can be measured between them. We 
denote this abstract point set by $M$. By the way the set $M$ was 
constructed, it is naturally endowed with the structure of 
\emph{metric space} \cite{Nash}. The \emph{distance} between a given 
pair of points refers to an observation of lengths in multiples of 
some reference object called the \emph{rigid body}. (Observe here 
that numerical values of distances depend on the chosen rigid body 
but distances can be discussed without a specified reference body.) 
Furthermore, the metric induces the structure of \emph{topological 
  space} for $M$ \cite{Nash}. The dimension of $M$, which is two 
or three, must be the same at every point of space $M$.

\subsection{Manifolds}

For computational analysis, we need to characterize 
\emph{displacements} arithmetically and thus must 
\emph{parameterize} the point set. In the classical 
view this is done with coordinate systems, which assign unique 
\emph{coordinates}---that is, a tuple of real numbers---to each 
point of $M$. Formally this means a mapping of type 
$M \rightarrow \mathbb{R}^n$. Furthermore, $\mathbb{R}^n$ is 
endowed with the norm topology and the mapping is homeomorphic, 
i.e. continuous bijection with continuous inverse. However, the 
abstract point set $M$ is often ignored completely and the points 
of $M$ are identified with the coordinate numbers. Furthermore, 
often a metric connotation is given to the coordinate numbers. 
Thus the domain of the BVP is taken to be some particular subset 
of $\mathbb{R}^n$. Yet the coordinates are just labels for the 
points of $M$, such that they allow \emph{arithmetic} and need
not carry metric information. The coordinates can be changed 
with diffeomorphic mappings of type 
$\mathbb{R}^n \rightarrow \mathbb{R}^n$. 

Instead of a particular coordinate system, the modern view uses 
manifolds as domains for BVPs. In manifolds, the abstract point 
set $M$ is the primary object, which is not
identified with any particular set of coordinate numbers. The 
coordinate systems, or \emph{charts}, are considered secondary, 
which is natural because we can choose the charts in many ways, 
and the real numbers assigned to each point are not canonical. 
In particular, \emph{the physics---work and energy---does not 
  depend at all on the chosen chart}. Furthermore, $M$ is not
expected to be coverable by a single chart, as was the case for 
coordinate systems: Charts are homeomorphic mappings from the 
open connected subsets of $M$ to the open subsets of $\mathbb{R}^n$. 
Moreover, every point of $M$ is required to be in the domain 
of some chart. Any topological space, that can be covered with 
charts, is called a \emph{topological manifold} \cite{Boothby}. 
Thus the structure of topological manifold emphasizes the 
\emph{existence of charts}, rather than a particular chart.

Topological manifold is not enough for physics, because the 
differentiability of functions needs an extra structure. Let 
us require a collection of charts, that are bound together by 
their \emph{transition maps}, or change-of-chart maps: The 
\emph{differentiability} of a real-valued function $h$ defined 
on $M$ is defined indirectly through the differentiability of the 
functions $h\circ f^{-1}$, where $f$ is a chart. The chart can be 
changed with the transition map $f\circ g^{-1}$ as follows: 
$(h\circ f^{-1})\circ(f\circ g^{-1}) = h\circ g^{-1}$. We do not 
want the $n$-times differentiability of $h$ to depend on the choice 
of chart, and therefore require that the transition map be (at least) 
$n$-times differentiable. If this holds, the charts are considered 
equivalent. This relation constitutes an equivalence class, 
a \emph{differentiable structure}, of charts. The set $M$ together 
with a differentiable structure is a \emph{differentiable manifold} 
or here just \emph{manifold} for short 
\cite{Boothby}\cite{Jänich}\cite{Warner}.

\subsection{Standard Parameterization}

 \begin{figure}
  \centering\includegraphics[width=0.9\textwidth]{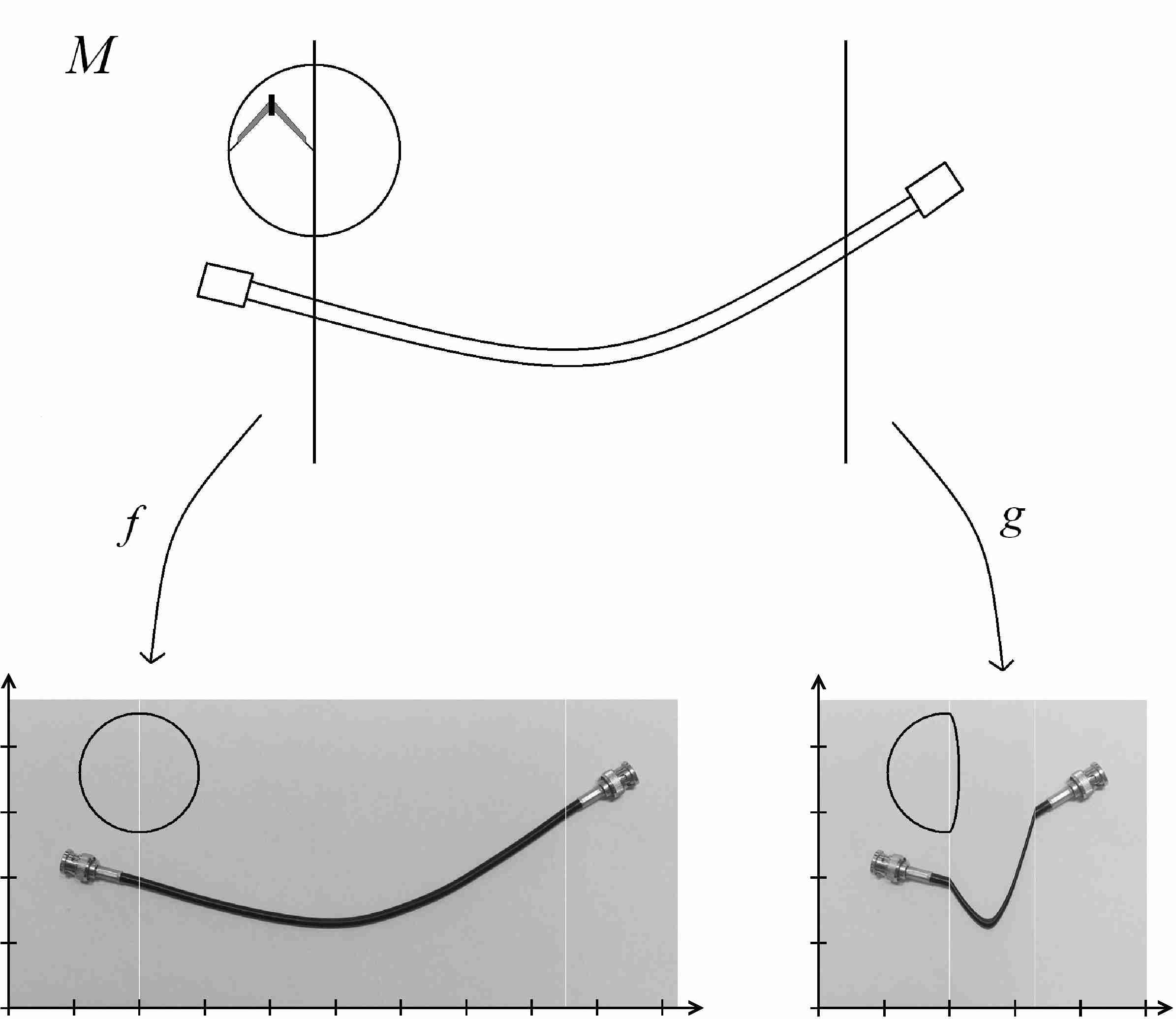}
  \caption{Standard parameterization: 
    The line drawing on top refers to a real coaxial cable.  
    The rigid body we chose gave us a pair of dividers and enabled 
    us to specify spheres. Chart $f$ from $M$ to $\mathbb{R}^n$ is 
    a standard parameterization, for the images of the spheres in 
    $M$ are also spheres in $\mathbb{R}^n$ in the sense of Euclidean 
    2-norm. Consequently, chart $g$ is not a standard parameterization.} 
  \label{fig:charts} 
\end{figure}

\begin{figure}
  \centering\includegraphics[width=0.55\textwidth]{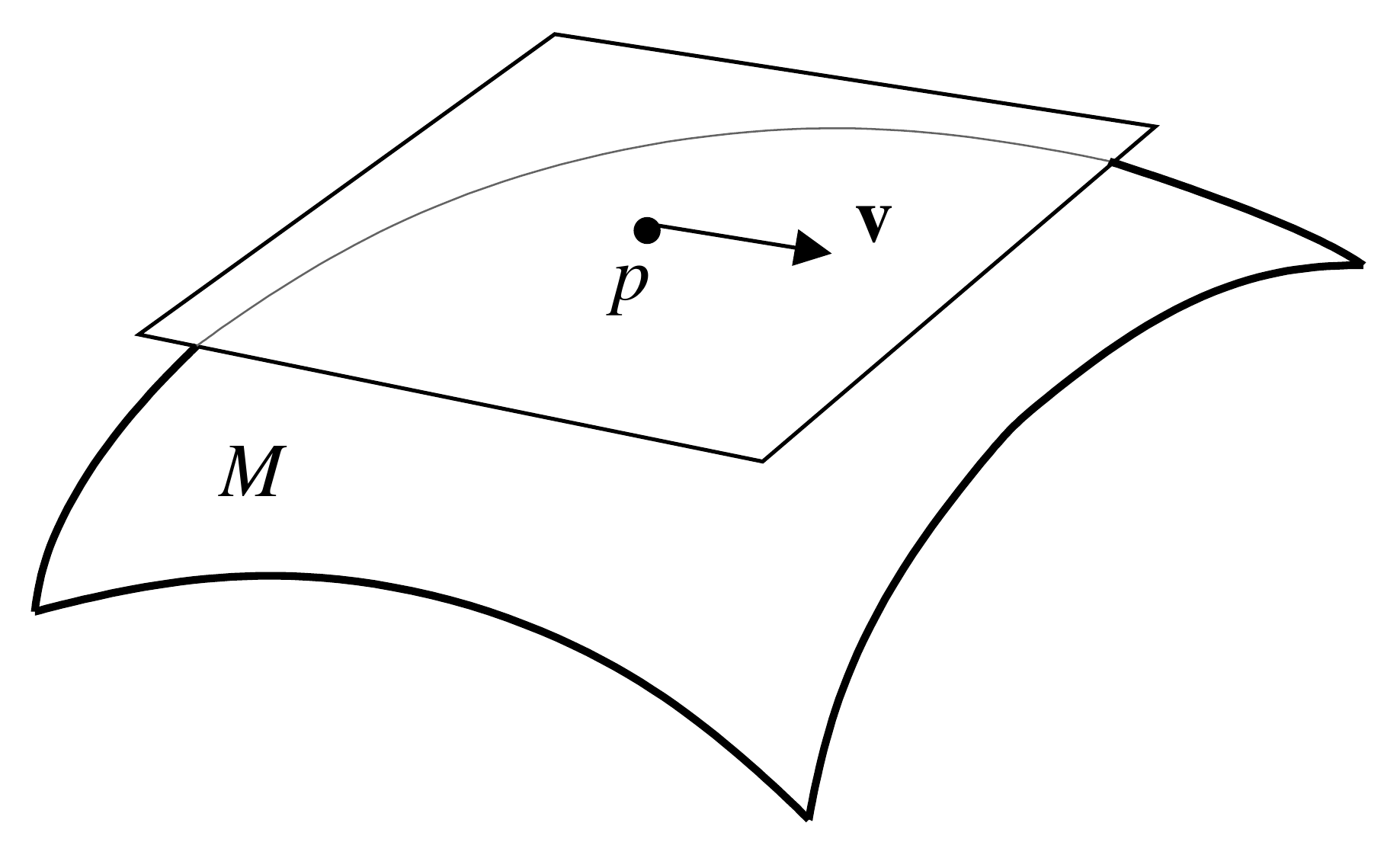}
  \caption{Tangent space and tangent vector. A tangent vector 
    $\mab{v}$ of the tangent space of the point $p$ in manifold $M$.} 
  \label{fig:tangent} 
\end{figure}

The rigid body can be used as a pair of dividers to specify spheres 
in $M$. If the image of every sphere in $M$ under a chart is a sphere 
also in $\mathbb{R}^n$ when measured with the {\em Euclidean 2-norm}, 
we call the chart a \emph{standard parameterization}. For example, 
the chart $f$ in figure~\ref{fig:charts} is a standard parameterization. 
All standard parameterizations are identical up to scaling, rotation, 
and translation (plus reflection). The standard parameterizations are 
important because in literature the numerical values of the material 
parameters are given in a specific class of standard parameterizations. 
This is natural because the numerical values of the material parameters 
are meaningful only with respect to a given metric, and in practice, 
the given metric is always based on a length unit system in which 
some rigid body is chosen for reference.

\subsection{Tangent Spaces}

Charts enable us to talk about displacements in small neighborhoods, 
i.e. \emph{virtual translations}: the possibility to parameterize 
$M$ implies a local $n$-dimensional vector space at each 
point of $M$. The vector space is called the \emph{tangent space}, 
and the elements of this space are \emph{tangent vectors} 
\cite{Boothby}\cite{Jänich}\cite{Warner}\cite{Burke}. Virtual 
translations can be recognized as tangent vectors, such as 
$\mab{v}$ at point $p \in M$ in figure~\ref{fig:tangent}. Note 
here that whereas in the classical vector analysis the fields 
are mappings from the domain to a single vector space, the points 
of a manifold all have their distinct tangent spaces.

The \emph{electric field} is an object that yields the virtual 
electromotive force $dU$ (virtual work up to charge) for every 
virtual translation $\mab{v}$. In other words, the electric field 
is a functional field over the tangent spaces. Correspondingly,
the \emph{electric flux} is an object that yields the virtual 
flux for every pair of virtual translations, which define  
a virtual surface. That is, the electric flux is also a functional 
field over the tangent spaces. The solutions of quasi-static BVPs 
are such functional fields.

\subsection{Inner Product}

In the classical view we use a vector field to express the electric 
field. This possibility is due to \emph{Riesz representation theorem} 
\cite{BossavitCompuEM}\cite{Yosida}, that enables us to represent the 
electric field as a \emph{vector field}, once an \emph{inner product} 
is available. In the classical view the inner product is defined in 
the codomain vector space, but in the modern view each tangent space 
of $M$ has its own inner product. The pointwise representation of the 
electric field functional by a vector $\mab{E}$ requires an inner 
product into each of them. At each point, with 
the inner product denoted by ($\cdot$,$\cdot$), the virtual emf 
$dU(\mab{v})$ corresponding to the virtual translation vector 
$\mab{v}$ can be represented as $dU(\mab{v}) = (\mab{E},\mab{v})$. 
This representation \emph{makes the inner product an inherent part 
  of the classical view}: the vector $\mab{E}$ depends on the chosen 
inner product whereas $dU$ does not. When the tangent spaces of 
a manifold are equipped with an inner product which varies smoothly 
from point to point, the manifold is called a \emph{Riemannian manifold} 
\cite{Boothby}\cite{Jänich}\cite{Warner}. 
The inner products on the tangent spaces of $M$ are usually called 
\emph{metric tensors}. They induce a metric~\footnote{The metric tensors 
  on the tangent spaces of $M$ are
  selected such that the induced metric on $M$ is the same as the
  metric $M$ already has: the metric given by the measurements with
  some rigid body.} on 
$M$ and also bring about the notion of \emph{angles} between the 
tangent vectors \cite{Boothby}.

\begin{figure}
  \centering\includegraphics[width=0.8\textwidth]{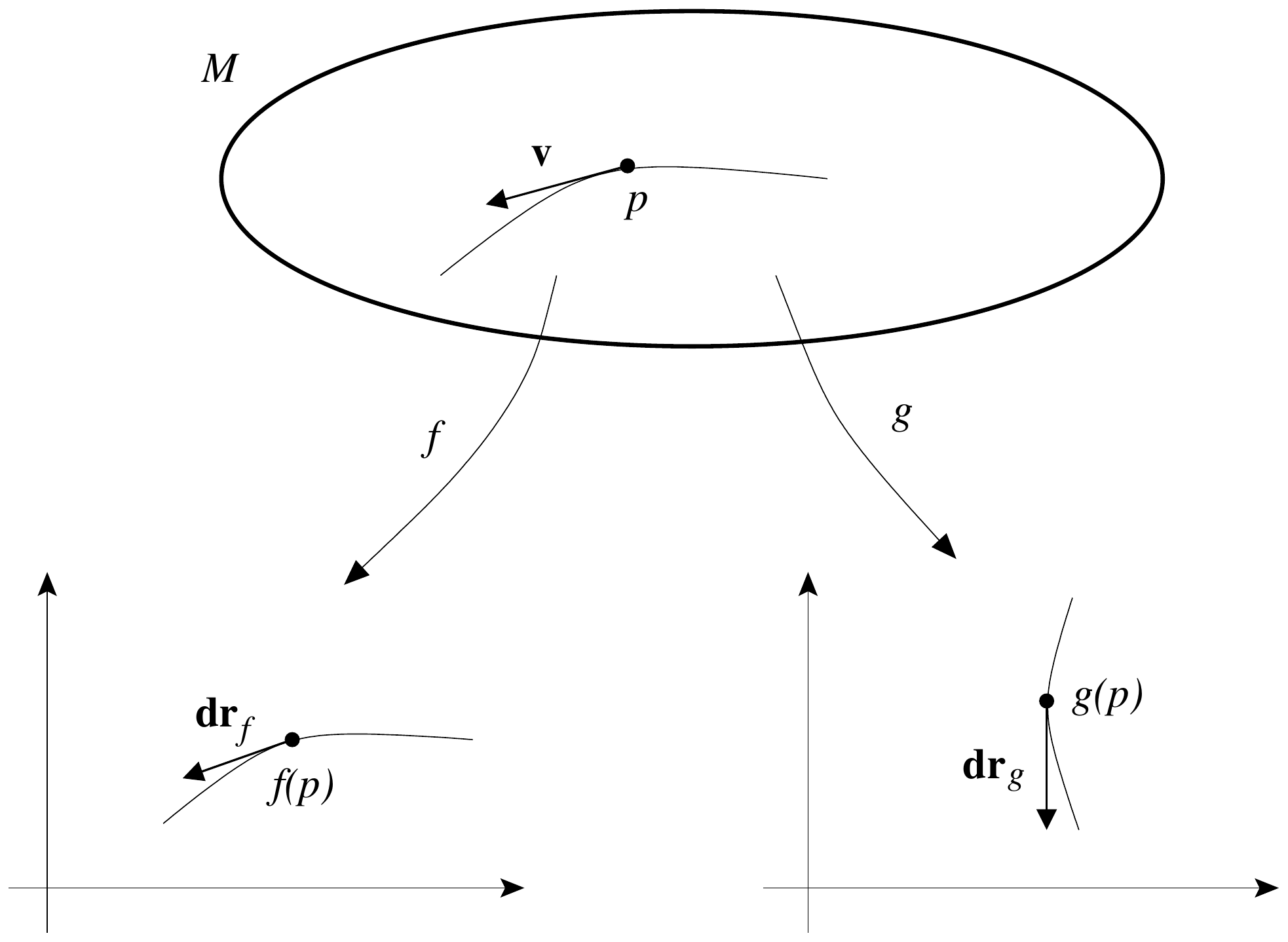}
  \caption{Correspondences of tangent vectors under different 
    charts. A tangent vector $\mab{v}$ at $p \in M$ corresponding 
    to vectors $\mab{dr}_f$ at $f(p)$ and $\mab{dr}_g$ at $g(p)$.} 
  \label{fig:images} 
\end{figure}

For practical computations with tangent vectors, such as evaluations 
of metric tensors, we need the counterparts of the tangent vectors 
and the metric tensors on ${\mathbb R}^n$. For each tangent vector 
$\mab{v}$ at $p$, a unique counterpart vector (the so-called 
\emph{push forward} \cite{Burke}) $\mab{dr}_f$ exists in 
${\mathbb R}^n$, induced by the chart $f$, see figure \ref{fig:images}. 
We can use these counterpart vectors $\mab{dr}_f$ to define 
metric tensors on ${\mathbb R}^n$ such that the tensors uniquely 
correspond to the ones on the tangent spaces of $M$: For each 
metric tensor $(\cdot,\cdot)_M$ given for $M$, there is 
a unique equivalent metric tensor $(\cdot,\cdot)_f$ for 
$\mathbb{R}^n$, that satisfies 
\begin{equation} 
  (\mab{dr}_f,\mab{dr}'_f)_f  \,=\, (\mab{v},\mab{v}')_M 
\end{equation} 
at all points and for all pairs $\mab{dr}_f$, $\mab{dr}'_f$ of 
$\mathbb{R}^n$ and all pairs of corresponding tangent vectors 
$\mab{v}$, $\mab{v}'$. Equivalent metric tensors on ${\mathbb R}^n$ 
and $M$ make $f$ an isometry. However, the metric tensor chosen
for $\mathbb{R}^n$ need not be equivalent with the one on $M$, 
i.e. the charts need not be isometries.

\section{Equivalent Descriptions of BVP}

The theory of electromagnetics is independent of the chosen metric 
tensor and the chart, as stated in the introduction. In fact, 
Maxwell's equations do not need a metric tensor at all. They are
meaningful on a differentiable manifold (not necessarily 
a Riemannian manifold) 
\cite{DiCarlo91}\cite{BossavitJapan98d}\cite{Bossavit_OnTheNot}: 
the virtual works related to the displacements of a point charge 
are independent of the metric tensors and chosen charts. Naturally, 
the energy stored in a system does not depend on the chosen charts 
either. However, the constitutive equations need to be described 
to define the energy and there are at least two ways to proceed: 
First, a metric tensor is selected in order to represent the 
fields as vector fields, and constitutive equations are defined 
couple the two vector fields. Second way to proceed is to couple 
the functional fields directly,in which case the constitutive 
equations couple different types of fields. Although a metric 
tensor is not needed to construct the constitutive equations in 
this case, constitutive equations induce metric tensors on the 
manifold \cite{Bossavit_OnTheNot}. Yet in neither strategy does 
the energy itself depend on the selected or the induced metric 
tensor. Thus an electromagnetic BVP is independent of particular 
chart or metric tensor, and it is uniquely specified with manifold 
$M$, Maxwell's equations, the constitutive equations, and the 
boundary values. 

However, the material parameters, or rather \emph{the numeric 
  values of permittivity, permeability, and conductivity depend 
  on the chosen metric}. Furthermore, to describe an abstract 
manifold concretely, we need to do that by charts, even 
though the manifold itself is independent of particular charts. 
Therefore, to pose the BVP, we need charts and metric tensors. 
When a chart and a metric tensor field are chosen, the numerical 
values of the material parameters can be specified. Then the 
material parameters are real- or matrix-valued fields defined 
on the chart. Thus we have a triplet \{chart, metric tensor, 
material parameters\} that describes a geometry of the BVP 
domain and the constitutive equation. Because we can choose any 
chart from the differentiable structure, and define any metric 
tensor on the chart, there are infinitely many geometries that 
we can use to pose the BVP. In fact, there is an equivalence class 
of these triplets \{chart, metric tensor, material parameters\},
that describe the same unique BVP. We call the equivalence 
relation producing these classes the \emph{material equivalence}.

We can now express a BVP on any chart: we need a triplet \{chart, 
metric tensor, material parameters\}, Maxwell's equations, and the 
boundary conditions. To express the BVP on any other chart, we only 
need to select another triplet, because Maxwell's equations are 
independent of the metric and of particular chart, as are the 
boundary conditions. Notice, that we can also keep the same chart 
and change the metric tensor and the material parameters. These 
descriptions of the BVP have equivalent solutions; that is, in 
any description, the electric vector field corresponds to the same 
virtual emf functional field. We obtain a new view of BVPs from the 
material equivalence: a BVP can be seen as a material equivalence 
class together with Maxwell's equations and boundary conditions. 
This view suggests an approach to construct solver software systems:
\begin{quote}
Solver for BVPs should allow the user to select any triplet from
the material equivalence class.
\end{quote}
This way the user can use the triplet that is the most suitable 
for numerical solution. 

The benefit of choosing any triplet is that we can solve 
numerically a larger class of problems. Let us derive the 
material parameters for given chart and metric tensor, such that
the material equivalence holds with another triplet that is known.

\section{Material Parameters for Given Chart And Metric Tensor}

To determine the material parameters for given chart and metric 
tensor, we require that the energy stored in a system and the 
virtual works related to the displacements of a point charge be 
invariant of our choice of chart and metric tensor. For simplicity, 
let us now focus on electrostatics, because other quasi-static 
cases can be treated conceptually in a similar fashion.

Following the standard convention, we use $\mab{dr} \cdot \mab{dr}'$ 
to denote $(\mab{dr},\mab{dr}')$ in $\mathbb{R}^n$. Also, for each 
metric tensor field $\cdot$ in $\mathbb{R}^n$, there is a unique 
symmetric positive definite matrix field $S$ such that 
$\mab{dr} \cdot \mab{dr}' = \mab{dr}^T S \, \mab{dr}'$ holds 
pointwise for all $\mab{dr}$ and $\mab{dr}'$. Let us assume we 
know a triplet \{$f_i$, $\cdot_i$, $\epsilon_i$\}, consisting of 
chart $f_i$ of $M$, metric tensor $\cdot_i$ on $\mathbb{R}^{n}$ 
and the material parameters $\epsilon_i$. Next, we choose chart 
$f_j$ from the differentiable structure and use metric tensor 
$\cdot_j$ on $\mathbb{R}^{n}$ and then determine the material 
parameters $\epsilon_j$ for the triplet 
\mbox{\{$f_j$, $\cdot_j$, $\epsilon_j$\}} such that the two triplets 
are equivalent.

\subsection{Invariance of Virtual Work}

The counterparts or the push forwards of a tangent vector $\mab{v}$ 
at $p \in M$ are $\mab{dr}_i$ and $\mab{dr}_j$ under charts $f_i$ 
and $f_j$, respectively. The virtual emf $dU(\mab{v})$ corresponding 
to the virtual translation $\mab{v}$ is $\mab{E}_i \cdot_i \mab{dr}_i$, 
or \mbox{$\mab{E}_j \cdot_j \mab{dr}_j$}, where $\cdot_i$ and 
$\cdot_j$ are metric tensors in $\mathbb{R}^{n}$. These expressions 
must equal pointwise:  
\begin{equation} 
  \label{eq:virtenergyequi}
  \mab{E}_j \cdot_j \mab{dr}_j = \mab{E}_i \cdot_i \mab{dr}_i. 
\end{equation} 
 
Let $f_j \circ f_i^{-1}: \mathrm{ran}(f_i) \subset \mathbb{R}^{n} 
\rightarrow \mathbb{R}^{n}$  
be the transition map from the range of $f_i$ to the codomain
of $f_j$. Denoting by $J$ the \emph{Jacobian matrix} \cite{Stromberg} 
of $f_j \circ f_i^{-1}$, we have 
\begin{equation} 
  \label{eq:equicoord} 
  \mab{dr}_j = J \, \mab{dr}_i. 
\end{equation} 
Substituting (\ref{eq:equicoord}) for $\mab{dr}_j$ in 
(\ref{eq:virtenergyequi}), and relying on the invariance of the 
virtual work, we get 
\begin{equation} 
  \label{eq:virtenergy} 
  \mab{E}_j^{T} S_j \, J \, \mab{dr}_i 
  = \mab{E}_i^{T} S_i \, \mab{dr}_i 
  \,\,\,\,\,\,\, \forall \,\, \mab{dr}_i, 
\end{equation} 
where $S_i$ and $S_j$ are the matrix presentations of the metric 
tensors $\cdot_i$ and $\cdot_j$. Because 
(\ref{eq:virtenergy}) holds for every point and for every virtual 
translation $\mab{dr}_i$, we can express $\mab{E}_i$ in terms of 
$\mab{E}_j$ and matrices $S_i$, $S_j$, and $J$ as follows: 
\begin{equation} 
  \label{eq:innewcoord} 
  \mab{E}_i = S_i^{-1} J^{T} S_j \, \mab{E}_j. 
\end{equation}
If we know either $\mab{E}_i$ or $\mab{E}_j$, we know $dU$ for every 
virtual translation, hence the solution of the BVP.

\subsection{Invariance of Energy}

The energy $W$ stored in the electric field in $f_i(M)$ can be 
expressed as  
\begin{equation} 
  \label{eq:energyequ} 
  W = \int_{f_i(M)} \mab{E}_i \cdot_i \epsilon_i \, \mab{E}_i \, dv_i 
  = \int_{f_i(M)} \mab{E}_i^{T} S_i \, \epsilon_i \, \mab{E}_i \, dv_i, 
\end{equation} 
where the matrix $\epsilon_i$ contains the material parameters in the
chart $f_i$. Equation (\ref{eq:energyequ}) induces an inner product 
\begin{equation} 
  \label{eq:energyinnerproduct} 
  ( \mab{E}_{k} , \mab{E}_{l} )  
  = \int_{f_i(M)} \mab{E}_{k}^{T} S_i \, \epsilon_i \, \mab{E}_{l} \, dv_i 
\end{equation} 
for the fields.
Here $\mab{E}_{k}$ and $\mab{E}_{l}$ are any two vector fields. 
Even though there is a metric tensor in the expression of the inner 
product of the fields, the energy is independent of chosen metric 
tensor: the product $S_i \epsilon_i$ describes the constitutive 
equation, such that the value of energy is the same regardless of 
the choice of $S_i$. This is what is meant by the claim that 
electromagnetic theory is independent of the chosen metric. 
Note that the inner product defined in (\ref{eq:energyinnerproduct}) 
allows us to formulate the Galerkin method. 

For any electric field, the energy stored in the field should be 
invariant under the choice of chart, or
\begin{equation} 
  \label{eq:energyequimatrix} 
  \int_{f_i(M)} \mab{E}_i^{T} S_i \, \epsilon_i \, \mab{E}_i \, dv_i  
  = \int_{f_j(M)} \mab{E}_j^{T} S_j \, \epsilon_j \, \mab{E}_j \, dv_j 
\end{equation} 
must hold for any equivalent fields $\mab{E}_{i}$ and $\mab{E}_{j}$. 
Using the \emph{change of variables theorem} \cite{Stromberg}, we 
can write the left hand side of (\ref{eq:energyequimatrix}) in the 
codomain of $f_j$ as 
\begin{eqnarray} 
  \int_{f_i(M)} \mab{E}_i^{T} S_i \, \epsilon_i \, 
  \mab{E}_i \, dv_i 
  = \int_{f_j(M)} \mab{E}_i^{T} S_i \, \epsilon_i \, 
  \mab{E}_i \Big\arrowvert J^{-1} \Big\arrowvert \, dv_j, 
  \label{eq:energymatrixchange} 
\end{eqnarray} 
where $\Big\arrowvert J^{-1} \Big\arrowvert$ is the determinant of the 
inverse of the Jacobian matrix $J$. Substituting the expression given in  
equation~(\ref{eq:innewcoord}) for $\mab{E}_i$ in 
(\ref{eq:energymatrixchange}), we get 
\begin{eqnarray} 
  \int_{f_j(M)} \mab{E}_j^{T} S_j \, J \, S_i^{-1} S_i \, \epsilon_i \, 
  S_i^{-1} J^{T} S_j 
  \mab{E}_j \Big\arrowvert J^{-1} \Big\arrowvert \, dv_i 
  = \int_{f_j(M)} \mab{E}_j^{T} S_j \, \epsilon_j \, \mab{E}_j \, dv_j. 
  \label{eq:energymatrixfinal} 
\end{eqnarray}
Now, because this holds for any vector field $\mab{E}_j$ defined 
on $f_j(M)$, the matrix $\epsilon_j$ can be written in terms of 
matrices $\epsilon_i$, $S_i$, $S_j$, and $J$, as follows: 
\begin{equation} 
  \label{eq:epsilon2} 
  \epsilon_j = J \, \epsilon_i \, S_i^{-1} J^{T} S_j 
  \Big\arrowvert J^{-1} \Big\arrowvert.  
\end{equation}

\section{Equivalence of Numerical Solutions}

\begin{figure}[!t]
  \centering\includegraphics[width=0.9\textwidth]{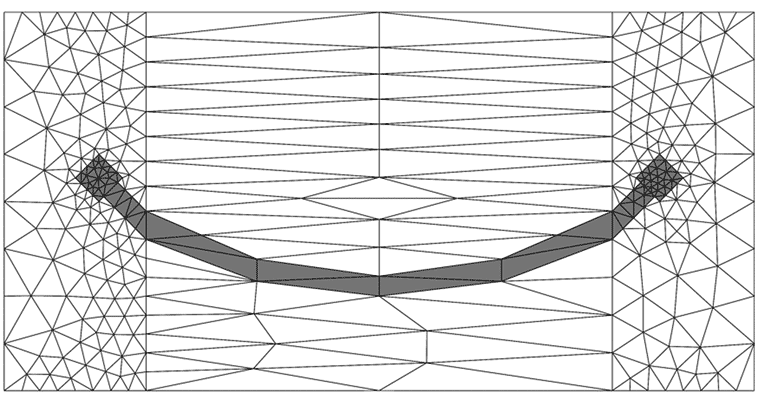}
  \includegraphics[width=0.45\textwidth]{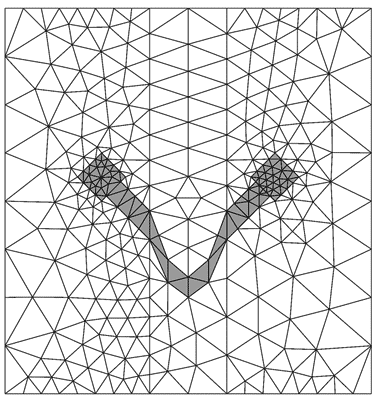}
  \caption{Example of equivalent meshes. Two equivalent meshes for 
    the codomains of the charts $f$ and $g$ of figure \ref{fig:charts}.
    The mesh in the codomain of the standard parameterization $f$
    has long elements in the middle, which may imply a poor quality.
    However, it is not always possible to generate any reasonable
    mesh for the standard parameterizations.}
  \label{fig:elements} 
\end{figure} 

The solutions for equivalent triplets are equivalent from 
a mathematical point of view, and so are the numerical solutions: 
Next we define equivalent meshes and show that under equivalent meshes, 
the system matrices in FEM are identical up to the round-off errors of 
the floating point representation.

First, assume a mesh in $M$. The mesh points in $M$ have corresponding 
points in every codomain of the chart of every triplet; consequently, 
we get an equivalent mesh for every chart (an example of equivalent 
meshes in figure~\ref{fig:elements})~\footnote{Notice that if the edges 
  of mesh elements are straight in some chart, in an equivalent mesh of 
  some other chart the edges can be curved.}~\footnote{Elements that are 
  long in one direction only can give good solutions if the field has 
  little variations in that direction.}.
Assuming real number arithmetics, equation (\ref{eq:energyequimatrix}) 
implies that the system matrices related to them are identical by 
construction. Thus the FEM solutions are also equivalent. However, 
in practice the system matrices are assembled and the numerical 
solutions are calculated with floating point numbers, and this leads 
to round-off errors. Thus the system matrices are identical only up 
to the round-off errors. 

Although the solutions are equivalent, in practice every BVP has 
triplets---sometimes including the standard parameterizations---that 
do not allow mesh generation, and thus no numerical solution. In these 
cases, mesh generation may well be possible with some other triplets. 
However, if the mesh quality criteria are not changed according to 
the chosen triplet, then there is usually a price to pay: the mesh 
quality may be poor, which causes numerical error in the solution. 
For example some elements may be very long in the standard 
parameterization as seen in top of figure~\ref{fig:elements}.

\section{Applications}

We now have all the necessary tools to express a BVP, attempt
a solution, and interpret it with any triplet, and next suggest 
how to take advantage of this possibility in practice. 
First, we show how a chart can be changed when the metric tensor 
in $\mathbb{R}^n$ is fixed, which is usually the case. Second, 
we suggest a scheme to use multiple charts in mesh generation
and problem solution. Third application is open boundary problems
which are solved by selecting a triplet in which the problem
domain in the chart is bounded. Finally, we discuss how to model
motion with only one mesh by changing the metric.

\subsection{Reparameterization Under Fixed Chart Metric}

Because the numerical values of the material parameters are always 
given in literature in some length unit system, it is easiest
to describe the constitutive relations with isometric standard 
parameterizations. Consequently, the programmer of a finite element 
software system often assumes that the user applies only isometric 
standard parameterizations; that is, the user can control only the 
unit of length, and rotate and translate the model. Furthermore, due 
to the isometry assumption, the metric tensor is implicitly fixed 
such that it induces the Euclidean 2-norm. Therefore the matrix 
representing the metric tensor in the intrinsic coordinate system 
of the solver is implicitly hardwired to be the identity matrix. 

Let us assume that we know how to express a BVP in some isometric 
standard parameterization $f$, that is, we know the triplet \{$f$, 
$S_f$, $\epsilon_f$\}. However, instead of using $f$ directly, we 
would like to reparameterize with chart $g$, which is not a standard 
parameterization, to avoid floating point problems in numerical 
modeling. Because of the same hardwired metric tensor in the 
codomains of $f$ and $g$, the distances between any two coordinates 
(tuples of real numbers) do not change when changing chart from 
$f$ to $g$. However, a change of chart makes the coordinates 
correspond with different points in the manifold $M$, and thus 
the change of chart in this case implies a change of metric, hence 
the chart $g$ is not isometry.

To set up our BVP correctly under $g$, we must impose the material 
equivalence on $g$. For this we can use only the material parameters 
$\epsilon_g$, because $g$ is chosen by the modeller, and the metric 
tensor is hardwired by the programmer. Equation (\ref{eq:epsilon2}) 
shows that the material parameters $\epsilon_g$ require information 
on $g$ relative to $f$. Yet everything related to $f$, including $f$ 
itself, exists only in the modeller's mind. We are solving 
the BVP using a triplet \mbox{\{$g$, $S_g$, $\epsilon_g$\}}, and we 
interpret this description of the BVP under $g$ as being equivalent 
to that under $f$. However, the software programmer who assumes only 
standard parameterizations and has no information about $f$, might 
think of $g$ as a standard parameterization and very likely interpret 
the problem geometry very differently. In this sense, we would be 
``misusing'' the software. 
 
To impose the material equivalence, we need the Jacobian between 
the charts and material parameters $\epsilon_f$. After substituting 
$I$ for $S_f$ and $S_g$ in (\ref{eq:epsilon2}), we get the 
following equivalent material parameters: 
\begin{equation} 
  \label{eq:epsilon2new} 
  \epsilon_g = 
  \frac{1} {\Big\arrowvert J \Big\arrowvert} J \, \epsilon_f \, J^{T},  
\end{equation} 
where the identity  
$\Big\arrowvert J \Big\arrowvert 
\Big\arrowvert J^{-1} \Big\arrowvert = 1$ 
is used (Henrotte et al. \cite{Henrotte_Finite} arrived at a similar 
expression but one of their motivations was open boundary problems, 
see section 6.3). It is easy to see from (\ref{eq:epsilon2new}) that 
the material parameters in $g$ may not be a scalar multiple of the 
identity matrix, even if they were in the chart $f$, a point 
expressed in \cite{Bossavit_OnTheNot} from the opposite point of view.   
 
The solution of the problem are fields $\mab{E}_g$ and $\mab{D}_g$. 
However, to present the fields in the standard parameterization $f$, 
where they may be more intuitive, we must use (\ref{eq:innewcoord}) 
and $\mab{D}_f = \epsilon_f \, \mab{E}_f$ to obtain the equivalent 
fields.  
 
\begin{remark} $\mab{D}_g$ cannot be transformed into $\mab{D}_f$ 
  with the same transformation as $\mab{E}_g$ is transformed into 
  $\mab{E}_f$ in (\ref{eq:innewcoord}). That is, the electric field 
  intensity $\mab{E}$ and the electric flux density $\mab{D}$ 
  transform differently between charts. To get the $\mab{D}_f$, 
  one can use (\ref{eq:innewcoord}) to transform $\mab{E}_g$ to 
  $\mab{E}_f$ and then use the constitutive law 
  $\mab{D}_f = \epsilon_f \, \mab{E}_f$.
\end{remark} 
 
The above method is fully realizable with any mesh generator and 
solver software that allows for pointwise definition of material 
parameters as a matrix. However, if the solver does not allow for 
pointwise definition, only charts $g$ with piecewise affine 
transition maps from the standard parameterization $f$ can be used.

\subsection{Modelling With Multiple Charts}

\begin{figure}
%  \centering\includegraphics[width=0.9\textwidth]{fig6_multiple.eps}
  \centering\includegraphics[width=0.9\textwidth]{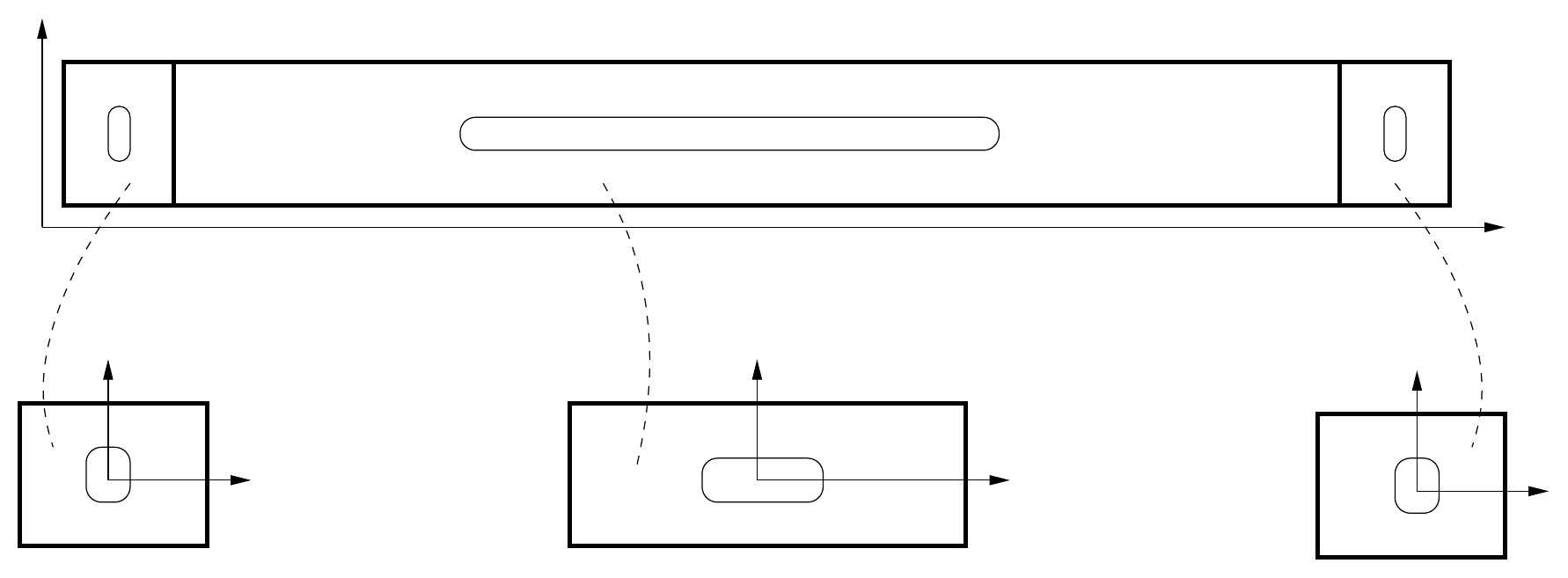}
  \caption{Multiple charts. Top: Standard parameterization, with 
    three rectangular regions. Bottom: Three charts, that each 
    cover one of the rectangular regions, but the origins are
    moved and scales are changed.}  
  \label{fig:multiple} 
\end{figure}

Previously we had two charts that both cover the whole domain. 
However, manifold can be covered with multiple charts, such that 
only a part of the manifold is covered by each chart. All that 
is required is that every point of the manifold belongs to the 
domain of some chart and that all the charts belong to the same 
differentiable structure. The use of multiple charts can benefit 
numerical analysis, for example in the case of 
figure~\ref{fig:multiple}: First of all, the mesh generation gets
easier, because the detailed regions could be covered separately 
with their own charts, with the origins located such that there 
are as many floating point numbers available as possible. Second, 
for the same reason, the calculations of the inner products for 
the system matrix are more accurate.

Even though multiple charts are used, we still assume here that 
the manifold \emph{can} be covered with one $\mathbb{R}^n$ chart. 
This makes it possible to partition the manifold into separate 
domains such that the domains only overlap on their boundary. 
Furthermore, this guarantees that the manifold can be embedded 
into the $n$-dimensional Euclidean space. Next, based on these 
assumptions, we suggest an implementation of multiple charts into 
software systems.

First, the user implicitly decides on one isometric standard 
parameterization, which covers the whole domain. Let us call 
it the universal chart, because it works as the reference to 
the other charts, and the material parameters are given on it. 
However, the universal chart is not used in the calculation or 
in the mesh generation, but these are done in separate charts 
that cover only parts of the manifold: the user partitions the 
universal chart into multiple regions, such that the regions 
only overlap on their boundary (see figure~\ref{fig:multiple}). 
Then the user gives new charts for these regions and gives the 
transition maps to the universal chart. In practice the user 
only gives the separate charts and the transition maps to the 
software system, and then the universal chart is constructed 
from these by the software. It is also possible that the 
universal chart is not constructed at all in the software 
system but exists only implicitly. The metric tensors for 
these separate charts could be fixed or given by the user. 
The material parameters for the charts are given as if they 
are standard parameterizations, and the software calculates 
automatically the material parameters to be used in the 
calculations. This is possible because the metric tensors are 
known, as are the transition maps, whose Jacobians can be 
automatically calculated. 

Finally, in order to attain compatibility of meshes in different 
charts, the meshes have to agree at the boundaries of the regions. 
The mesh generation could proceed as follows: First a mesh is 
generated on one of the regions with the chart that covers it. 
Then the coordinates of the nodes on the common boundary with 
a second region are mapped with the transition map to the chart 
that covers the second region, and the rest of the mesh in the 
second region can be generated. The assembly of the system matrix 
can be done as usually.

\subsection{Open Boundary Problems}

\begin{figure}[t]
%  \centering\includegraphics[width=\textwidth]{fig7_open.eps}
  \centering\includegraphics[width=\textwidth]{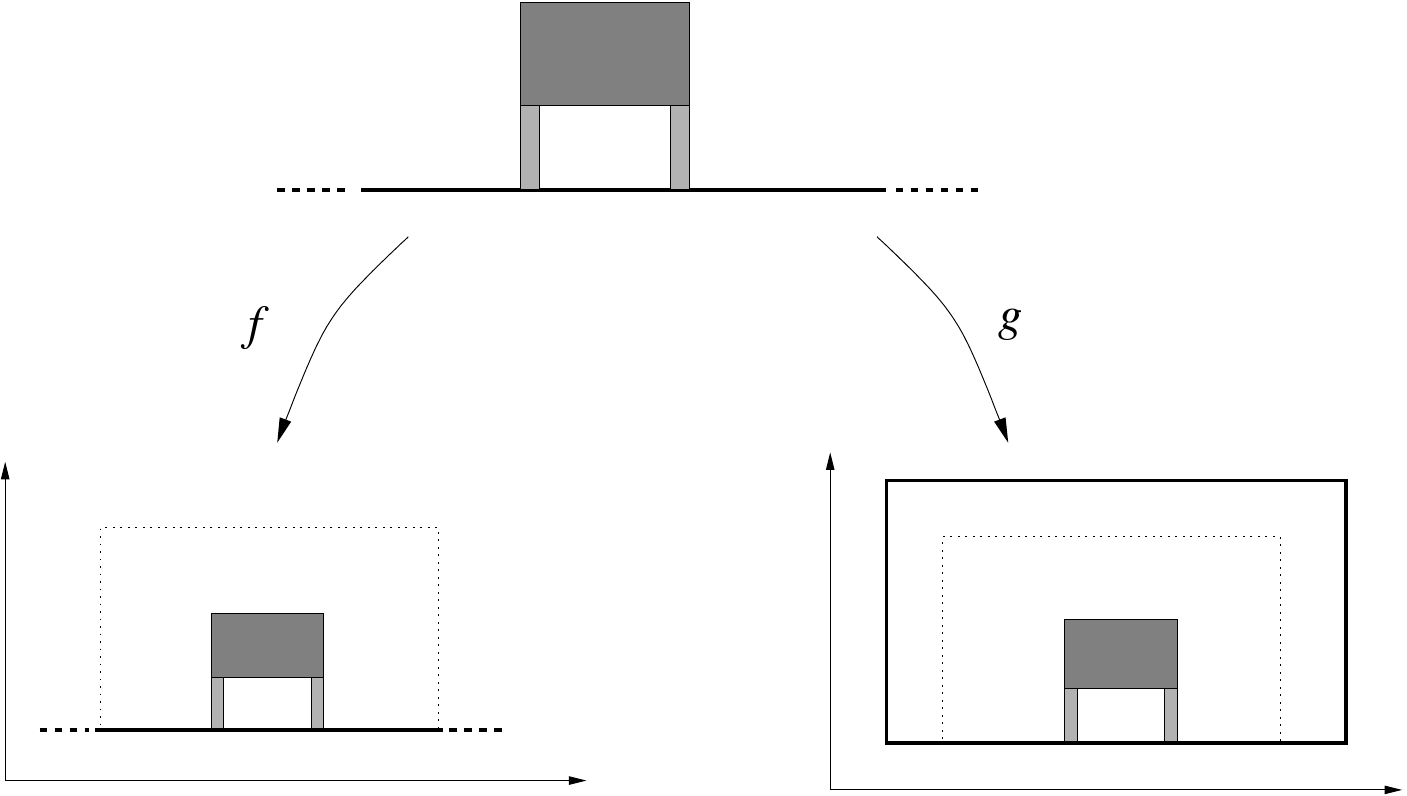}
  \caption{Open boundary problem. Top: The problem domain (manifold)
    consist of an device over the ground. Bottom: The codomains of 
    two charts, $f$ and $g$. The regions inside of the rectangles 
    denoted with dashed line correspond to the same part of the 
    domain, but in $g$ the rest of the space is between the 
    rectangles.}  
  \label{fig:open} 
\end{figure}

The open boundary problems are often solved by truncating the 
domain and setting the fields zero at an artificial distant boundary. 
This usually gives a good enough approximation for the fields near 
the sources, because the fields tend to zero quite fast as the 
distance from the source increases. The distance of the artificial 
boundary from the sources and regions of importance is decided by the 
modeller. However, the error from this truncation is hard to estimate 
and the number of elements needed to cover all that empty space can
be large.

To overcome these problems, many methods have been proposed over the 
years: The so called "ballooning method" \cite{Silvester}, in which 
the true distance of the boundary is put very far away with a thin 
layer of special elements. Another method uses infinite elements 
\cite{Bettess}, which are special decaying basis functions used 
in FEM. Yet another method couples the FEM with analytical 
solutions as in \cite{Silvester2}. Finally, the transformation 
(or shell transformation) methods presented for example in 
\cite{Henrotte_Finite}\cite{Imhoff} put the infinite boundary to 
a finite distance with the help of a suitable change of coordinates. 
However, many of these methods are hard to implement into most
production solvers, because they require modification of the solver 
codes: For example, the modeller needs to define new basis functions 
or needs to give pointwise Jacobians of the change of coordinates. 

The triplet approach of BVPs suggest a strategy to solve open 
boundary problems: Just select a suitable triplet, such that in 
the chart the infinity boundary is at a finite distance, such as 
the chart $g$ in the figure~\ref{fig:open}. This is effectively 
identical to many of the above methods. For example, in the 
transformations method the Jacobian of the change of coordinates 
is implemented in the software code whereas in the triplet approach 
the Jacobian is included in the material parameters given by the 
user.

\begin{figure}
%  \centering\includegraphics[width=\textwidth]{fig8_mapping.eps}
  \centering\includegraphics[width=\textwidth]{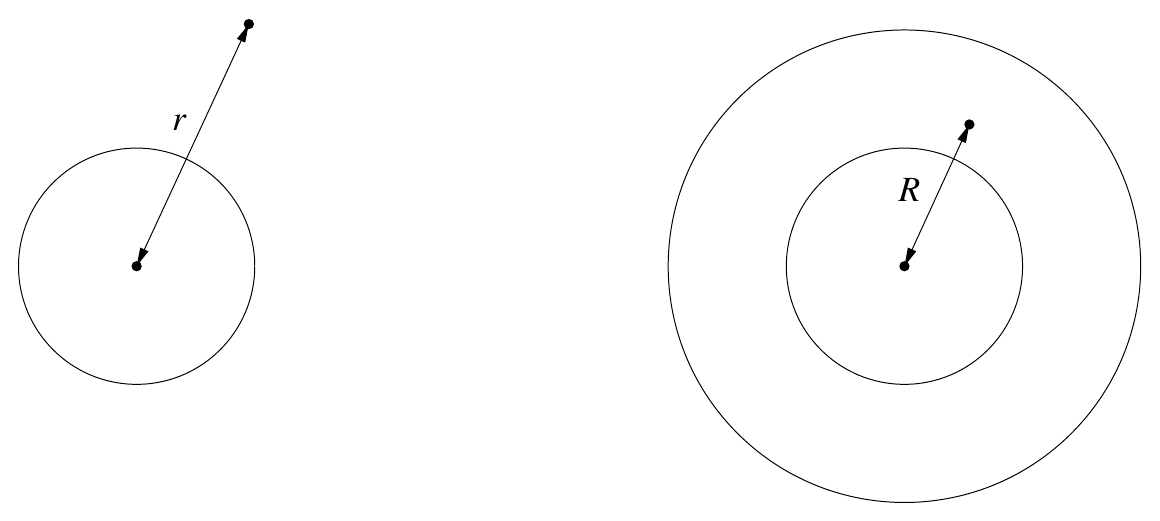}
  \caption{Mapping of infinite domain to finite domain. The region
    outside of the unit disc in the left hand side of the figure 
    is mapped to ring of radius two in the right hand side of the 
    figure. A point at distance $r$ from the center of unit disc
    is mapped to a point at distance $R$ from the center.}  
  \label{fig:mapping} 
\end{figure}

The triplet should be selected such that a small region including 
the source regions and other regions of interest is modelled as 
usually, but the complement of this region is mapped to a small
bounded outer region as in the chart $g$ in the 
figure~\ref{fig:open}. The elements nearest the infinity boundary 
of the domain are such that the points at the boundary are mapped
to the infinity by the transition map form $g$ to the corresponding
standard parameterization. A suitable transition mapping could be 
the following: The region outside the unit disc is mapped to the 
ring with exterior radius two, such that the ring surrounds the 
unit disc and the boundary of the disc is the inner boundary of 
the ring, see figure~\ref{fig:mapping}. A point outside the unit 
disc at the distance $r$ from the center of the unit disc is mapped 
to a point at the distance $R$ from the center, such that the 
distances have the following relation:
\begin{displaymath}
  R = 2 - \frac{1}{r}.
\end{displaymath}

This method can be used with almost any solver: Only the material 
parameters have to be changed accordingly, which is a responsibility 
of the modeller. However, the procedure is easy to automate: the 
user only specifies the regions where the outer boundary is mapped
to the infinity by the transition map from $g$ to the
standard parameterization, and the software system would take 
care of the rest. The system could have built-in such a mapping or 
mappings which adapt according to the coordinates of the infinity 
boundary and the boundary of the infinite regions. If necessary,
the user gives the correspondence between these boundaries and then 
the system modifies the user-given material parameters according 
to the adapted mapping.

\subsection{Modelling Motion With Single Chart}

\begin{figure}
%  \centering\includegraphics[width=\textwidth]{fig9_movement.eps}
  \centering\includegraphics[width=\textwidth]{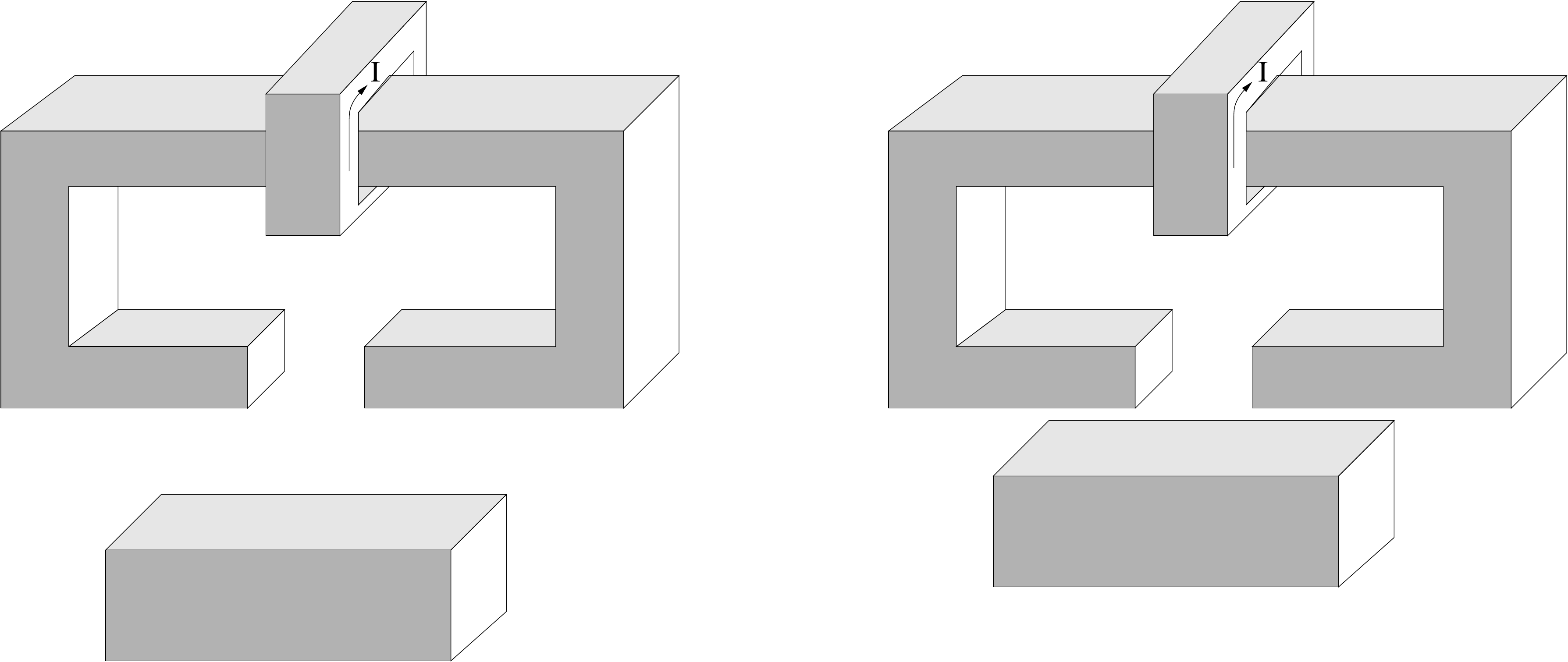}
  \caption{Modelling motion. Electromagnet: coil with current $I$, 
    ferromagnetic core and load, which is attracted upwards.}  
  \label{fig:movement} 
\end{figure}

When modelling motion, we have to solve a sequence of BVPs whose
only difference is that some object changes its position relative 
to the other objects. For an example, consider the force calculations 
of an electromagnet, see figure~\ref{fig:movement}: The magnet 
attracts the load, which then moves upwards. One solution is to 
make whole model for each position of the moving object. For each 
position, we have to define a new geometry and generate a new mesh 
for it, and furthermore assembly the whole system matrix again. 
This is very time-consuming. 

The triplet approach can speed up the process: Keep the chart and 
the material parameters fixed and change the metric tensor. This 
changes the geometry and takes into account the motion: When 
a rigid object moves in a fluid (air) such that the topology does 
not change, i.e. the moving object does not touch any other material 
than the fluid, then the underlying differentiable manifold of the 
domain does not change (this underlying manifold is often referred 
to as material manifold). However, the \emph{Riemannian manifold} 
will change, because the metric tensor changes. This means that the 
distances between points of the manifold are changed. Thus we may 
select a single chart $g$ from the differentiable structure, to 
cover all the different Riemannian manifolds.

The metric tensor for the selected chart $g$ again depends on 
the material equivalence. This requires that some standard 
parameterization $f$ for each Riemannian manifold is known, 
when the corresponding Jacobians are also known. Then the matrix 
$S_g$ representing the metric tensor on $g$ can be solved from 
equation~(\ref{eq:epsilon2}): Assume the material parameter 
$\epsilon_f$ to be a scalar and require the material parameters 
to be equal, that is $\epsilon_g = \epsilon_f$. Then $S_g$ in 
terms of the Jacobian can be solved:
\begin{equation} 
  \label{eq:metrictensor} 
  S_g = \Big\arrowvert J \Big\arrowvert J^{-T} J^{-1}.  
\end{equation} 

Now, for the example in the case of figure \ref{fig:movement}, the 
metric tensor only changes in the air between the magnet and the 
load according to the equation (\ref{eq:metrictensor}). Because 
the mesh remains the same, only the elements in the system matrix 
corresponding to the mesh elements in this air region will change.
Thus only a partial re-assembly of system matrix is needed for each 
new step. Furthermore, in the case of iterative solvers, 
a preconditioner used for one step can be used effectively for many 
successive steps, because the changes in the system matrix can 
be very small. Also the previous solution may be a good initial 
guess. Thus, all these things---only one mesh, partial re-assembly 
of system matrix, the preconditioner, and the initial guess---can 
speed up the solution process substantially.

Many solvers do not offer the possibility to change the metric 
tensor, yet the modelling of motion with a single mesh is still 
possible: we can equivalently change the material parameters, see 
section 6.1. One needs to know a standard parameterization for 
each Riemannian manifold and then use the material parameters 
given by the equation~(\ref{eq:epsilon2new}). Finally, all things 
said about modelling motion apply also for modelling the geometric 
effects of deformations, such as the effects of magnetostriction.

\begin{figure}[t]
%  \centerline{\includegraphics[width=0.9\textwidth]{fig10_chart_stan.eps}}
  \centerline{\includegraphics[width=0.9\textwidth]{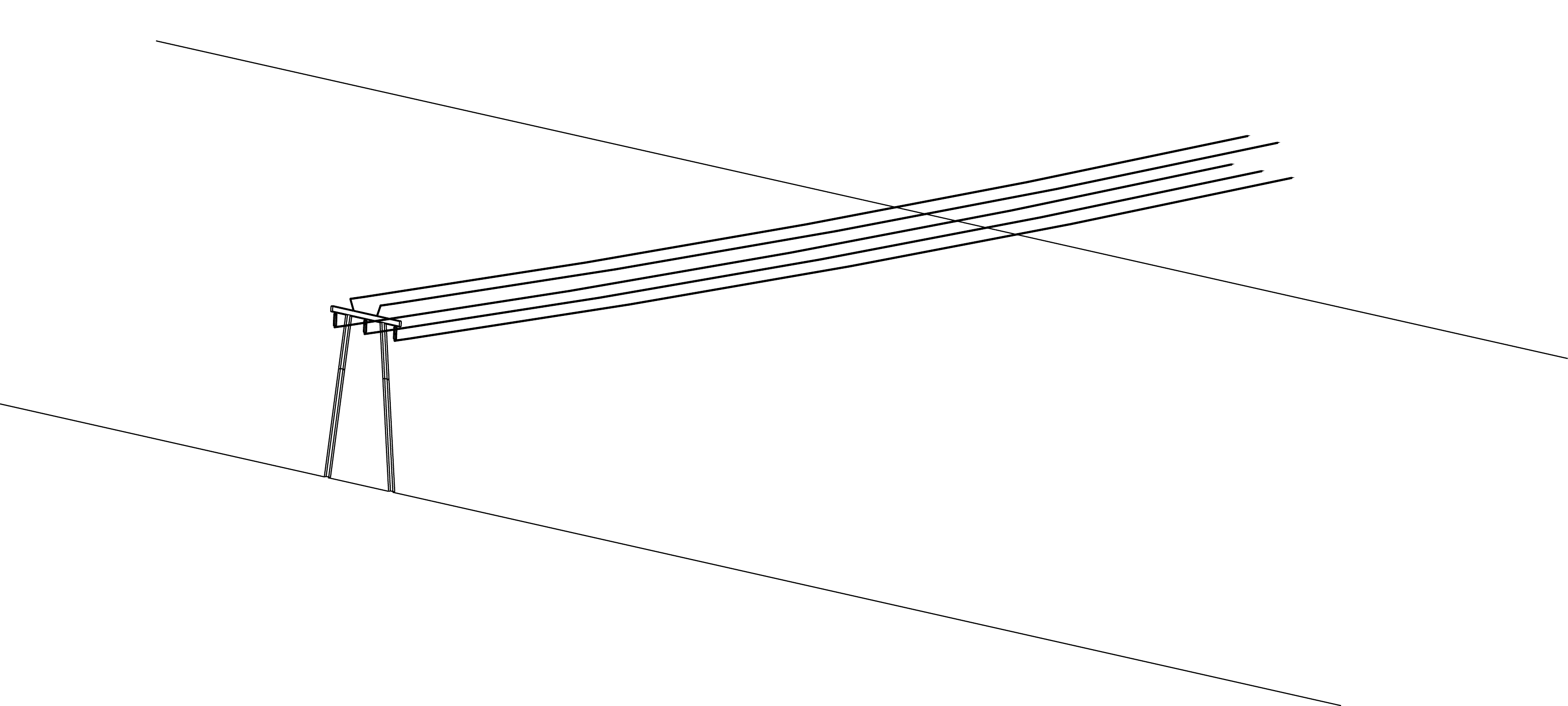}}
  \caption{Power line in a standard parameterization.}
  \label{fig:geom_stan}
\end{figure}

\begin{figure}[!h]
%  \centerline{\includegraphics[width=0.9\textwidth]{fig11_chart.eps}}
  \centerline{\includegraphics[width=0.9\textwidth]{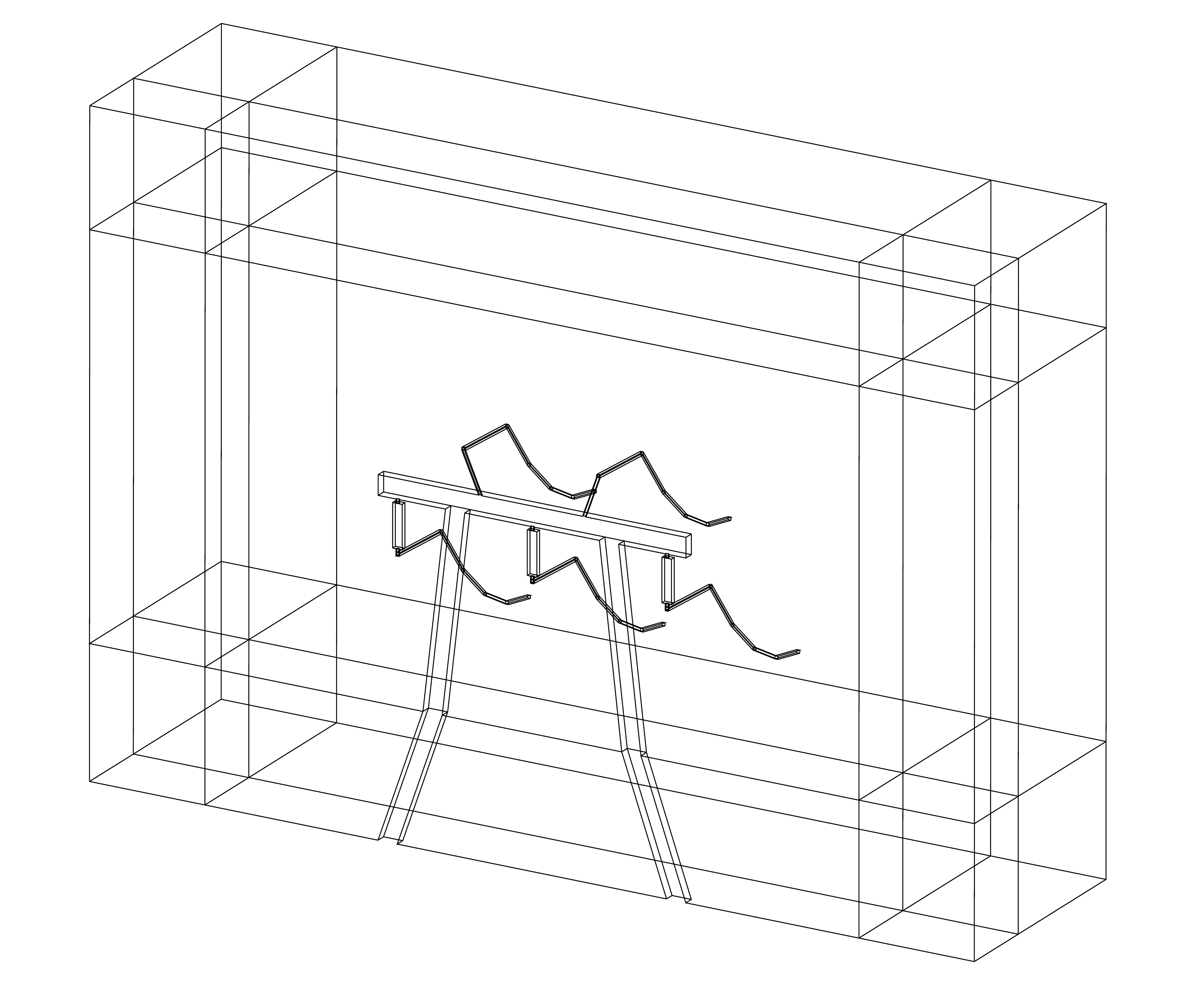}}
  \caption{The chart used in the power line problem.}
  \label{fig:geom}
\end{figure}

\begin{figure}[t]
  \centerline{\includegraphics[width=\textwidth]{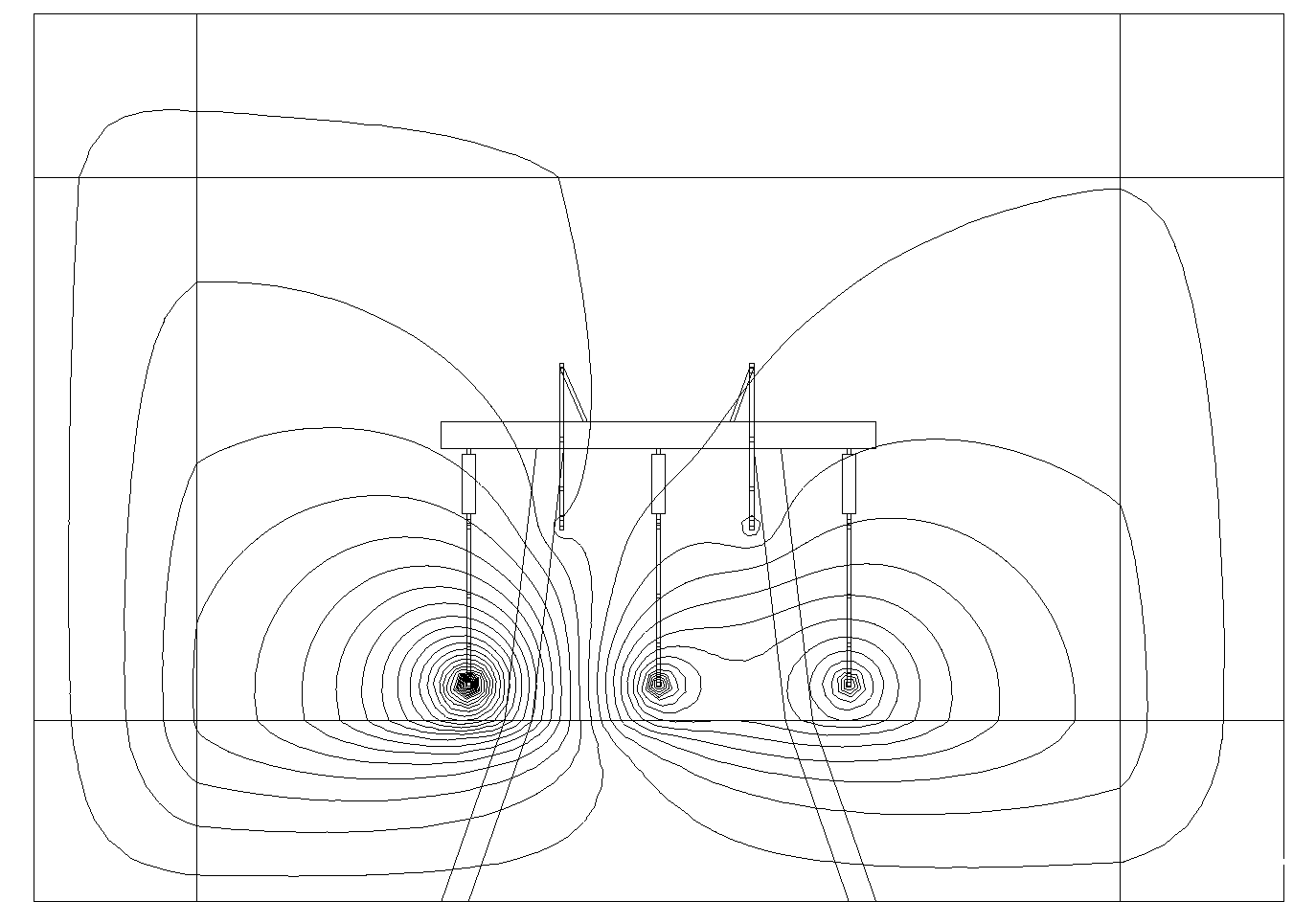}}
%  \centerline{\includegraphics[width=\textwidth]{fig12_result.eps}}
  \caption{Isovalue lines of potential of the power line 
    shown in the chart used in the calculations.}
  \label{fig:potentials}
\end{figure}

\section{Numerical Example}

We demonstrate a combination of proposed applications: an open 
boundary problem (section 6.3) with reparameterization under fixed 
chart metric (section 6.1). The problem to solve is the electric 
potential of a three-phase high-voltage line, a 3D-Laplace's 
problem. Figure~\ref{fig:geom_stan} shows a standard 
parameterization of the domain. The domain is the half-space 
above the ground apart from the line itself. The infinity boundary 
is mapped to a finite distance from the line, as explained in 
section 6.3. There is also great variation in the scale of the 
details of the domain: the length of the lines is order of hundreds 
of meters, whereas the smallest dimensions are order of centimeters. 
Based on experience, we can say that the mesh generation with the 
standard parameterization is difficult or even impossible. To 
avoid mesh generation problems, the lines are scaled down, as 
is the height of the pillars. The model used in the calculations 
is shown in figure~\ref{fig:geom}. The result is shown in 
figure~\ref{fig:potentials}. The figures and the calculations 
are produced with GetDP~\cite{getdp} and Gmsh~\cite{gmsh}.

\section{Conclusion}

Most solver software systems for quasi-static electromagnetic 
boundary value problems are based on the formulation of 
electromagnetics with classical vector analysis. This has many 
restrictions for numerical solution, because many mathematical 
structures have limited freedom or they are completely fixed. 
For example, the fields are presented as mappings from some 
global coordinate system to a single vector space and the 
metrics for the coordinate systems and the vector space are 
fixed. The restrictions limit the possible coordinate systems 
that can be used. 

The formulation of electromagnetics with manifolds and 
differential geometry is less restrictive and thus can help 
in a numerical solution. This view exposes the independence of 
the electromagnetic theory on the chosen coordinates or metric. 
Based on this it is shown that there is an equivalence class of 
triplets \{coordinate system, metric, material parameters\} for 
each boundary value problem. A triplet describes a geometry of 
the problem domain and the constitutive equations which are 
needed to pose the problem in the coordinate system. We propose 
that the solvers should allow to use any triplet, that is, the 
modeller should be able to choose any coordinate system and 
metric at will. These possibilities come built-in in 
differential geometry, whereas they are not self-evident in 
classical vector analysis. We discuss how this can be done. 
It is not hard to apply some of the triplet strategies in most 
production solvers, because not all of them require 
modifications to the code. We show that despite the freedom in 
the choice of the triplet, the FEM system matrices of a given 
BVP, assembled for a mesh represented under different triplets, 
are identical. 

We show how the triplet approach can be exploited in 
applications: Potentially prohibitive difficulties of mesh 
generation in problem domains that have large scale variation 
can be alleviated by a suitable choice of coordinate system. 
This extends the class of solvable problems, although possibly 
at the expense of mesh quality. Open boundary problems can be 
solved by choosing a convenient triplet for the solution. 
Solution of problems involving motion and deformations can be 
accelerated when a single mesh is used on every time step 
and only the metric is altered.

\end{document}